\documentclass[times,preprint,review,nopreprintline]{elsarticle}

\usepackage{subcaption}

\usepackage{amssymb}
\usepackage{latexsym}
\usepackage{svg}

\usepackage{mathtools}

\usepackage{amsmath}
\usepackage{comment}
\usepackage{inputenc}

\usepackage[hidelinks]{hyperref}
\hypersetup{hidelinks}

\usepackage{tabularx}
\usepackage{booktabs}
\usepackage{subcaption}

\DeclarePairedDelimiter\floor{\lfloor}{\rfloor}

\definecolor{newcolor}{rgb}{.8,.349,.1}

\begin{document}

\title{VertDetect: Fully End-to-End 3D Vertebral Instance Segmentation Model}

\author[1]{Geoff Klein}
\author[2,4]{Michael Hardisty\corref{cor1}}
\author[2,3,4]{Cari Whyne}
\author[1,2]{Anne L. Martel}

\address[1]{Department of Medical Biophysics, University of Toronto, Toronto, Canada}
\address[2]{Physical Sciences, Sunnybrook Research Institute, Toronto, Canada}
\address[3]{Department of Biomedical Engineering, University of Toronto, Toronto, Canada}
\address[4]{Department of Surgery, University of Toronto, Toronto, Canada}

\cortext[cor1]{Senior author credit is mutually shared between Michael Hardisty(m.hardisty@utoronto.ca) and Anne Martel(a.martel@utoronto.ca)}

% \begin{frontmatter}

\begin{abstract}
Vertebral detection and segmentation are critical steps for treatment planning in spine surgery and radiation therapy. Accurate identification and segmentation are complicated in imaging that does not include the full spine, in cases with variations in anatomy (T13 and/or L6 vertebrae), and in the presence of fracture or hardware. This paper proposes \textit{VertDetect}, a fully automated end-to-end 3D vertebral instance segmentation Convolutional Neural Network (CNN) model to predict vertebral level labels and segmentations for all vertebrae present in a CT scan. The utilization of a shared CNN backbone provides the detection and segmentation branches of the network with feature maps containing both spinal and vertebral level information. A Graph Convolutional Network (GCN) layer is used to improve vertebral labelling by using the known structure of the spine. This model achieved a Dice Similarity Coefficient (DSC) of 0.883 (95\% CI, 0.843-0.906) and 0.882 (95\% CI, 0.835-0.909) in the VerSe 2019 and 0.868 (95\% CI, 0.834-0.890) and 0.869 (95\% CI, 0.832-0.891) in the VerSe 2020 public and hidden test sets, respectively. This model achieved state-of-the-art performance for an end-to-end architecture, whose design facilitates the extraction of features that can be subsequently used for downstream tasks. 
\end{abstract}

\maketitle

% \begin{keyword}
% Vertebrae Detection \sep Vertebrae Segmentation \sep 3D Detection; 3D Medical Segmentation
% \end{keyword}

% \end{frontmatter}

\section{Introduction}

Detecting and segmenting vertebrae in medical images, including computed tomography (CT) scans, is necessary for many clinical tasks including treatment planning, surgical intervention and radiation treatment \cite{Fourney2011,Sahgal2013a,Thibault2017}. Minimally invasive surgical procedures require robust and accurate vertebrae identification and segmentation. Intra-operative imaging requires fast methods of identification and segmentation of vertebrae.

Labelling vertebrae in medical images can be a laborious manual process requiring anatomical expertise to compare vertebral locations to other anatomical landmarks (ie, rib locations). Medical images may not always include the full spine and fracture or collapsed vertebrae can pose additional challenges to accurate labelling. Although the human spine typically contains 33 vertebrae consisting of (7 cervical, 12 thoracic, 5 lumbar, 5 fused sacral, and 4 fused coccyx bones) variations are not uncommon. Sacralization occurs when the L5 vertebra is fused to the sacrum resulting in what looks like a missing vertebra. The opposite, lumbarization, occurs when the S1 detaches from the fused sacral vertebrae resulting in what looks like an additional vertebra referred to as L6. Sacralization and lumbarization have an occurrence of approximately 5\% and 3\%, respectively \cite{Doo2020}. Thoracolumbar transitional vertebrae, known as T13, are also possible with an occurrence of approximately 11\% \cite{Doo2020}. 

Segmenting individual vertebrae can be challenging as neighbouring vertebrae are in contact at the facet joints. This task is further complicated due to variable fields-of-view (where all vertebrae are not necessarily imaged in a given scan) and variations in scan quality (differing slice thickness and resolution). Disease and trauma can also affect vertebral appearance (osteoporosis lowering bone density, tumour involvement changing bone deposition patterns, and fractures disrupting bone distribution and geometry). The presence of surgical implants can obscure bone morphology, disease, and potential fractures.

The automation of vertebral detection and segmentation have been widely explored. Statistical shape models \cite{Castro-Mateos2015,Klinder2009,Benjelloun2011a}, deformable fences \cite{Kim2009} and deformable atlases \cite{Hardisty2007a} have been used towards automating vertebral segmentation. Hardisty et al. \cite{Hardisty2007a} developed a semi-automated 3D vertebral body segmentation model for CT scans using deformable atlas based registration. Neubert et al. \cite{Neubert2012a} segmented vertebral bodies from 3D magnetic resonance (MR) imaging using a two-stage active shape model where the spine was first localized from the scan and approximate vertebral body positions were found using active rectangles. Vertebral bodies were then segmented using deformable statistical shape models. These methods exploited the fact that vertebrae share similar physical characteristics, but they require some type of initialization (usually in the form of manual fiducial markers) and can be computationally expensive. Varying metrics are used between papers, but in general, these previous studies have DSC ranging from approximately 0.85 to 0.93 for vertebral segmentation. Detecting and labelling vertebrae have also been performed using support vector machine models \cite{Lootus2014a}, generalized hough Transform models \cite{Klinder2009}, and random classification forests \cite{Glocker2013b}. However, These results showed promise (3.3 mm \cite{Lootus2014a} and 11.5 mm \cite{Glocker2013b} error), but these detection models require significant prior knowledge of the spine and its characteristics.

More modern approaches for both vertebral detection and segmentation have employed Convolutional Neural Networks (CNN). Chen et al. \cite{Chen2015} realized that rather than relying on low-level hand-crafted features, neural networks could be used that take advantage of high-level feature representations of images. They also saw the benefits of CNN’s over feed-forward neural networks as CNN’s take better advantage of the spatial information in an image. Furthermore, the GPU implementation of neural networks allows fast training due to parallelization. Chen et al. used a combination of CNN and more classical machine learning approaches by first using a random forest to coarsely detect vertebrae in CT scans. This was followed by a CNN to further refine the vertebral detection and uses a shape model to incorporate features of neighbouring vertebrae. 

Semantic segmentation was further improved with the U-Net architecture by Ronneberger et al. \cite{Ronneberger2015} which resulted in significant advancements in image segmentation, especially in the medical space and is heavily used. Kuok et al. \cite{Kuok2018} developed a U-Net model with skip connections to segment 2D CT axial slices of vertebrae. Klein et al. \cite{Klein2020} developed a similar model, using a 3D U-Net to segment the vertebral body from 3D CT scans. Both models required cropping a single vertebra at a time and could be used in conjunction with known vertebrae of interest or coupled with detection and localization models. Lessmann et al. \cite{Lessmann2018} went further to segment and label all vertebrae in 3D CT scans by iteratively segmenting different patches of the 3D scan using a U-Net and keeping track of previously detected vertebrae by using memory instance layers. 

Further work in vertebral detection has come from Zhao et al. \cite{Zhao2021} where a Faster-RCNN \cite{Ren2017b} like model was developed to detect vertebrae in 2D sagittal MR slices and message passing was used to share information between neighbouring vertebrae to improve the detection. Both Yang et al. \cite{Yang2017} and Cui et al. \cite{Cui2021} automatically detect vertebrae in 3D CT scans using Gaussian heatmap predictions from an encoder-decoder architecture. Yang et al. used message passing of the heatmap predictions to share information between neighbouring predictions, whereas Cui et al. used shape and spatial encoding features to get accurate anatomical labels.

Being able to both detect and segment vertebrae has been investigated by Cheng et al. \cite{Cheng2021b} who used cascading Dense U-Net models on both 2D slices and full 3D CT scans. The first Dense-U-Net model determined the centroid of each vertebra in a 2D axial slice. The predicted centroids were then used in a 3D Dense-U-Net to perform 3D segmentation on each vertebra. Altini et al. \cite{Altini2021b} combined both CNN and classical machine learning with k-Means and k-NN clustering to both detect and segment vertebrae. A CNN first performs semantic segmentation on each vertebra using a V-Net \cite{Milletari2016a}. Vertebral detection is then achieved using a semi-automated approach where semantic segmentations are processed, and centroids are determined in an iterative slice extraction tool. The user is required to specify the number of segmented vertebrae and the anatomical label of the top-most vertebra, as well as select the best slice for each vertebra in the sagittal plane. Segmentation is then performed using a k-NN classifier and the centroid chosen locations. The authors reported a DSC of 0.909 on a subset of the VerSe 2020 test set(50/113).

The VerSe segmentation and detection challenges (both 2019 and 2020) \cite{Sekuboyina2021,Sekuboyina2020,Liebl2021b} provided further advancements in CT vertebra detection and segmentation with the availability of a large open dataset of 3D CT scans with segmentation, vertebral body centroids, and class labels. Both Payer et al. \cite{Payer2020,Sekuboyina2021}, winner of the VerSe 2019 challenge, and Chen et al. \cite{Sekuboyina2021}, winner of the VerSe 2020 challenge (average dice similarity coefficients of 0.898 and 0.912 for Payer and Chen on the hidden test set, respectively), used a combination of cascading models to detect and segment all vertebrae in a 3D CT scan. Payer et al. used three cascading models where the first was a U-Net to isolate the spine in the larger CT scan. This was followed up by the second model which used a combination of U-Nets to determine the shape and position information of each vertebra to properly detect the vertebral body centroids of each vertebra. The final model was a U-Net which semantically segmented the vertebra by cropping the regions from the CT scans using the predicted vertebral body centroids. Chen et al. used a slightly different combination of cascading models for the 2020 VerSe challenge. The first model was a U-Net similar to Payer et al. to isolate the spine from the whole CT scan. This was followed by a second U-Net inspired by Lessmann et al. where vertebrae were iteratively semantically segmented. The final model was a 3D ResNet-50 \cite{Hea} to classify the semantic segmentations using both the predicted segmentations from the second model and the input CT volume. Chen et al. also employed a Deep Reasoning module \cite{Chen2020c} to ensure previous predictions were anatomically realistic. It is also worth noting that Payer et al. used a similar configuration in the VerSe 2020 challenge as well, placing second, but implemented a post-processing method after the second model to correct mislabelled centroids. 

More recent vertebral detection and segmentation networks have utilized transformer networks. Tao et al. \cite{Tao2022} developed a two-stage framework for vertebrae detection and segmentation on the 2019 VerSe dataset. They developed Spine-Transformer to detect the centroid of each vertebral body in a 3D CT scan and used these centroid detections in a secondary network for vertebrae segmentation. You et al. \cite{You2022,You2023} developed a single transformer network for detection and segmentation on the 2020 VerSe dataset. However, due to the computational expense of transformers, Tao et al. \cite{Tao2022} required the inputs to the transformers to be patches of the overall 3D CT image. The use of patches means that the model cannot obtain information on the whole spine at once and overall contextual information is lost. You et al. \cite{You2022,You2023} tried to alleviate this by using both patches and the full unpatched image with two transformers to capture more global context. However, this method required manual cropping of the full CT images before they could be used in the global transformer.

Previous work has shown promising results with iterative, patch-based, cascading models, semi-automated approaches, as well as 2D methods to detect and/or segment the vertebrae in a 3D CT scan. Multi-model approaches do not make use of shared feature representation for the different sub-tasks (ie, using the same feature maps for classification and segmentation). Using the full 3D input the shared information between all vertebrae and relevant anatomical landmarks (ie, ribs) can be leveraged at the same time for efficient labelling. This is not achievable in iterative/patch-based methods as the feature maps do not contain the same whole 3D information. However, multi-model approaches can also be less efficient as the sum of all parameters of the multi-model approaches can be greater than a single-model approach. Furthermore, the feature maps generated by a single end-to-end model that utilizes the full 3D inputs have the potential to be used for secondary clinically relevant prediction tasks, an example being downstream fracture prediction. Therefore, this work proposes a full end-to-end trainable model VertDetect that can process full 3D CT volumes for the spine, and extract features for vertebral detection and segmentation from the entire volume. This fully end-to-end model will provide a more efficient method to detect and segment vertebrae in 3D CT scans without relying on iterative, multi-model or patch-based techniques.

\section{Proposed Method}
This paper proposes a model to simultaneously segment and detect all vertebrae in a 3D CT scan in any field-of-view called \textit{VertDetect}. This architecture represents a fully end-to-end method that may be more computationally efficient than other cascading methods including multiple U-Net models and more recent transformer models. This is achieved by reusing the features from a common convolution backbone in both the detection and segmentation stages of the network, in a similar fashion to other instance segmentation models (Mask R-CNN \cite{He2017b}, RetinaNet \cite{Lin2020}, FCOS \cite{Tian2019,Tian2020}).

\section{Contributions}
In this paper, we present a full 3D end-to-end vertebral instance segmentation model which can detect the centroid and predict a bounding box for each vertebra, predict the correct anatomical label and segment the individual vertebrae.

Our specific contributions include the following:

\begin{itemize}

\item A novel architecture for VertDetect is presented inspired by the previous work of Yi et al. \cite{Yi2021}, Mask R-CNN \cite{He2017b} and FCOS \cite{Tian2019,Tian2020}.

\item Due to the model's ability to use full 3D volumes, VertDetect enables detection from arbitrary fields-of-view and when only partial spinal anatomy is presented within the CT scan.

\item Linear scheduling was tested during training for centroid detection to assist in convergence by combining a variant focal loss from CornerNet \cite{Law2020} with mean-square-error (MSE) loss. 

\item An initial training step, referred to as self-initialization, was tested with the idea of ensuring that centroid predictions have improved convergence and do not conflict with other loss functions. 

\item A Graph Convolutional Network (GCN) layer was tested to enable better classification and overall model stability by leveraging shared information and taking advantage of the known ordering of the vertebrae in the spine. 

\end{itemize}

\section{Methodology}
\subsection{VertDetect Model Architecture}

% \begin{figure}[h!]
\begin{figure*}[h!]
  \begin{center}
	\includegraphics[scale=0.6,width=\textwidth]{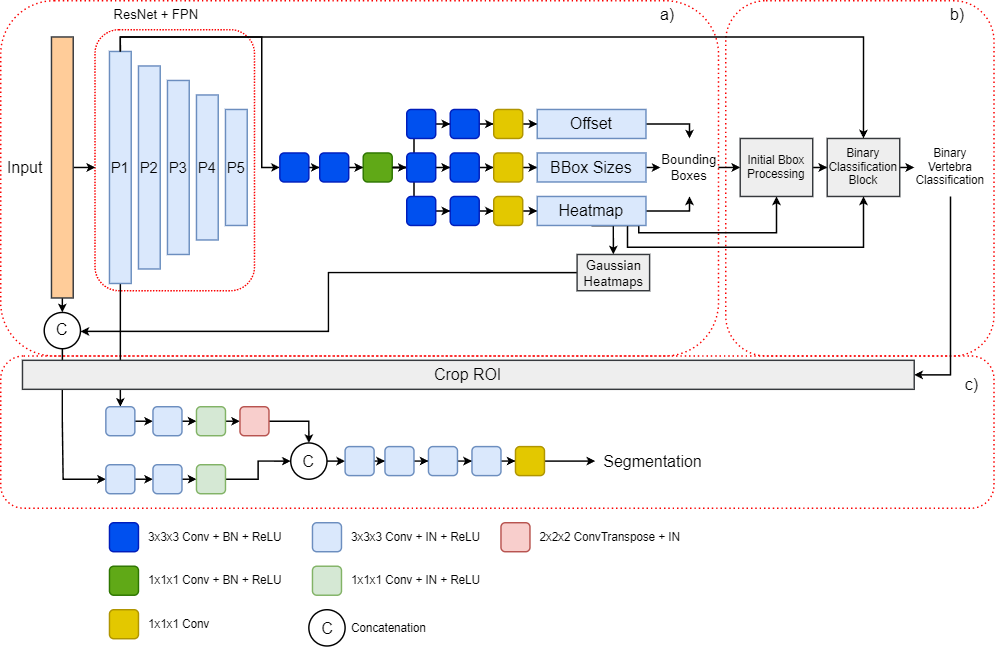}
	\caption{Block diagram for the VertDetect architecture. a) shows the Detection branch; b) the classification branch; c) the segmentation branch. The detection branch outputs the offset sizes, bounding box (Bbox) sizes, and heatmap. The classification branch determines what vertebrae exist in the CT image. The segmentation branch uses the outputs of the detection and classification branch to semantically segment positive candidates.} 
    \label{mod_arch}
    \end{center}
\end{figure*}

The proposed VertDetect model can be broken down into three main branches; detection, classification, and segmentation. The detection branch identifies each vertebra in a 3D CT scan by determining both its vertebral body centroid location and placing a bounding box around the whole vertebra. The classification branch utilizes shared information between each neighbouring vertebra to determine which vertebrae are present in the input CT scan. The segmentation branch semantically segments vertebrae that are detected from the classification and the detection branches.

\begin{figure}[h!]
    \centering
	\includegraphics[width=0.6\linewidth]{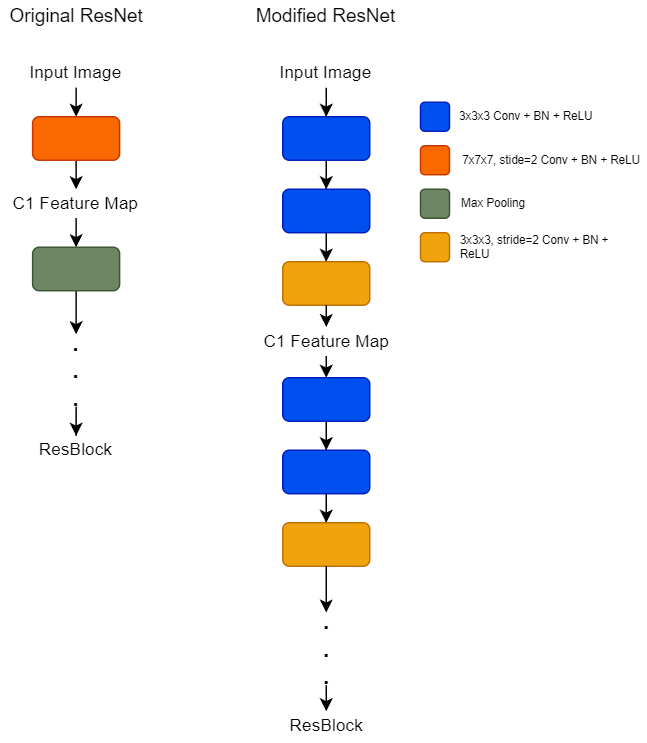}
	\caption{Modified ResNet-50 layers used in VertDetect.} 
    \label{resnet_mod}
\end{figure}

The overall architecture of VertDetect can be seen in Fig.~\ref{mod_arch}. A 3D ResNet-50 \cite{Hea} and Feature Pyramid Network (FPN) \cite{Lin2017} act as the backbone architecture. Feature maps from this backbone are then used in further downstream tasks. The FPN part of the backbone uses a consistent number of filters $d$. A modification to the original ResNet-50 architecture was implemented, and this can be seen in Fig.~\ref{resnet_mod}, which provides more information in the higher resolution feature maps.

\subsection{Detection Branch}

The detection branch utilizes an anchorless approach for object detection. It consists of three outputs: a heatmap predicting the vertebral body centroid location, an offset to the centroid locations to account for potential shifts due to downsampling, and bounding box sizes used to generate the bounding boxes centred on the vertebral body centroids. 

The largest resolution feature map from the convolution backbone (P1) is passed to three convolution layers, the first two having kernels of 3x3x3 and the last having a kernel of 1x1x1. The first of these three convolutions also compresses the P1 feature maps from $d$ to 128 to reduce the memory impact. The resulting feature map is then sent to three separate blocks of convolutions for generating the heatmap, calculating bounding box size, and determining the offset for each vertebra. Each block consists of two 3x3x3 convolutions followed by a 1x1x1 convolution with $C$, 3 and 6 channels for the heatmap, offset and bounding box sizes predictions, respectively, where $C$ is the number of potential vertebrae. All convolution operations, except for the final convolutions for the predictions, are followed by ReLU activations. Final predictions do not use activations.

The heatmaps provide a channel-wise centroid prediction where the maximum argument for each predicted channel corresponds to the location of the vertebral body centroid in a downsampled space. This downsampled space is consistent with the P1 output of the FPN (2x downsampling from input image size). To bring the downsampled centroid prediction to the full resolution, the offset sizes are used to fix potential misalignment of the original image and downsampled image caused by rounding errors that could have occurred when downsampling. The bounding box sizes then determine the size of the bounding box from the full resolution centroid. This allows for object detection to occur without the use of anchors. 

\subsubsection{Heatmap}

The heatmaps have C channels, where each channel corresponds to a vertebra (ie, channel 0 is C1, channel 1 is C2, etc.). Therefore, the overall classification of each centroid is implicitly defined in the channels of the heatmap predictions. The maxima of each channel's heatmap are used to provide the centroid predictions for each vertebra. Ground truth 3D Gaussian heatmaps are generated based on ground truth centroid points in a down-sampled space to match the size of the P1 output of the FPN (2x downsampling from input image size). The ground truth Gaussian distributions were constructed such that the peak max value was 1.0, $g(x,y,z) = e^{-\frac{(x-c_x )^2+(y-c_y )^2+(z-c_z )^2}{2 \sigma^2}}$, where $c_x$, $c_y$ and $c_z$ are the ground truth centroid coordinates, $\sigma$ is the standard deviation of the distribution (hyperparameter), and $x$, $y$, and $z$ are points in space.

In each channel of the heatmap, there is a single centroid point of interest and the rest is the background. To account for the large imbalance between foreground and background a variant focal loss was used as \cite{Lin2020}\cite{Law2020}\cite{Yi2021}:

\begin{equation}
      L_{focal} = -\frac{1}{N} \left.
\begin{cases}
(1 - p_i)^\alpha \text{log}(p_i), & \text{if } y_i = 1 \\
(1 - y_i)^\beta p_i^\alpha \text{log}(1 - p_i), & \text{else}

\end{cases}
   \right.
\end{equation}  

where $i$ is the $i^{th}$ index, $p$ is the predicted heatmap, $y$ is the ground truth, and $N$ is the number of centroids. This differs from the original focal loss \cite{Lin2020} by reducing the impact of predicted centroids that are close to the ground truth compared to predictions that are further when $y_i \epsilon [0, 1)$, with the $(1-y_i)^\beta$ term.

\begin{equation}
L_{heat} = aL_{focal} + bL_{MSE}    
\end{equation}

where $a=\frac{\epsilon}{\epsilon^\prime} \lambda$ and $b=\frac{\epsilon^\prime - \epsilon}{\epsilon^\prime}$ given some epoch $\epsilon$ and some threshold epoch $\epsilon^\prime$. This epoch threshold $\epsilon^\prime$ determines when to use the combined MSE and variant focal loss and when to switch to using only the variant focal loss. The constant $\lambda$ is a scaling term to address the large numerical difference between the variant focal loss and MSE functions and is $\lambda=\frac{1-\gamma}{\epsilon^\prime - 1}\epsilon^\prime + \frac{\gamma \epsilon^\prime - 1}{\epsilon^\prime - 1}$, and $\gamma$=1e-4. After epoch $\epsilon^\prime$, $L_{heat}=L_{focal}$. The overall heatmap loss is:

\begin{equation}
      L_{focal} = \left.
\begin{cases}
L_{MSE}, & \text{when } \epsilon = 0 \\
\frac{\epsilon}{\epsilon^\prime} \lambda L_{focal} + \frac{\epsilon^\prime - \epsilon}{\epsilon^\prime} L_{MSE}, & \text{when } 0 < \epsilon < \epsilon^\prime \\
L_{focal}, & \text{when } \epsilon \geq \epsilon^\prime 
\end{cases}
   \right.
\end{equation}

\begin{figure}[h!]
    \centering
	\includegraphics[width=0.5\textwidth]{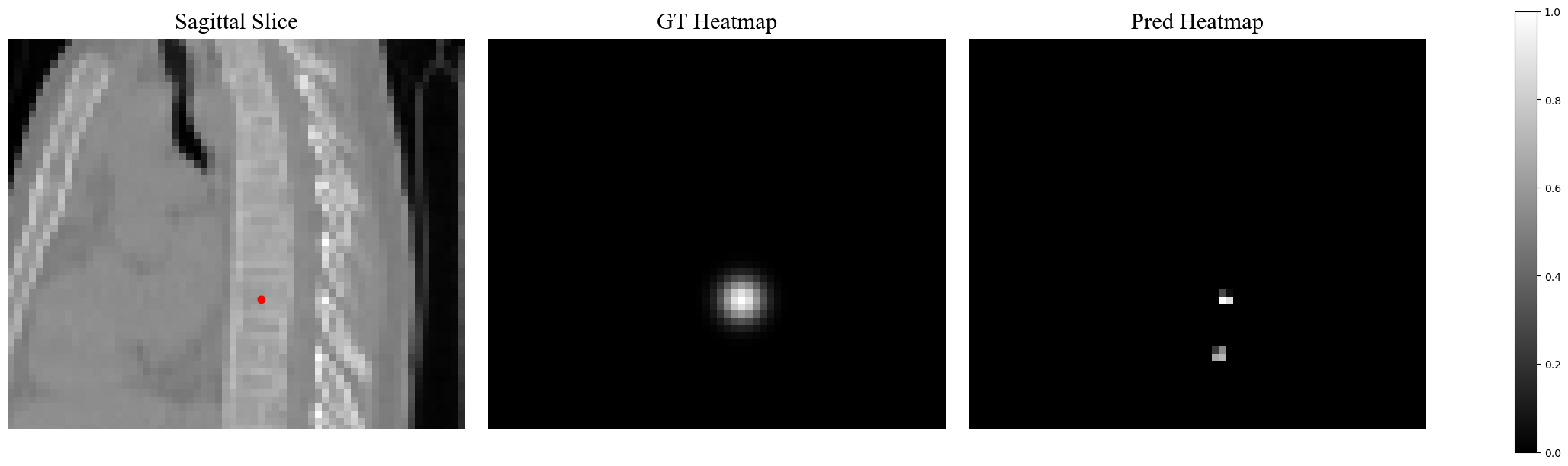}
	\caption{(left) Sagittal slice of CT image; (middle) ground truth heatmap for a single channel; (right) predicted heatmap for a single channel. The red dot in the sagittal slice corresponds to the ground truth centroid. The colour bar corresponds to the intensities of the predicted heatmap.} 
    \label{heatmap_slice}
\end{figure}

Fig.~\ref{heatmap_slice_better} shows an example of how predicted heatmaps result in small clusters. The local maxima of each cluster correspond to a vertebrae center. The maximum of the predicted heatmap corresponds to the predicted centroid location.

\subsubsection{Offset}

Following the work by Yi et al. \cite{Yi2021} an offset coordinate is used to shift the centroids to compensate for potential differences during upsampling:

\begin{equation}
o_i = \left( 
\frac{c_{x, i}}{n} -\floor*{ \frac{c_{x, i}}{n} }, 
\frac{c_{y, i}}{n} -\floor*{ \frac{c_{y, i}}{n} },
\frac{c_{z, i}}{n} -\floor*{ \frac{c_{z, i}}{n} } 
\right)
\end{equation}

where $i$ is the $i^{th}$ centroid, $c_{x,i}$, $c_{y,i}$, and $c_{z,i}$, are coordinates for the $i^{th}$ centroids. The brackets $\floor{}$ are the floor operation and $n$ is the downsampling size. A smooth-L1 loss is used to regress the offsets:

\begin{equation}
L_{offset} = \sum_{i} smooth_{L_1} \left( o_i - \hat{o_i} \right)
\end{equation}

where $o_i$ and  $\hat{o_i}$ are the ground truth and predicted, respectively.

\subsubsection{Bounding Box Sizes}

The bounding box sizes are used to determine the bounding box surrounding the centroid point. The coordinates of the bounding box for the $i^{th}$ vertebra ($bb_i$) is:

\begin{equation}
    \begin{array}{c c}

    \multicolumn{2}{c}{bb_i = \left[ x_0, x_1, y_0, y_1, z_0, z_1   \right] } \\
    x_0 = c_{x, i} - s_l, & x_1 = c_{x, i} + s_r \\ 
    y_0 = c_{y, i} - s_p, & y_1 = c_{y, i} + s_a \\ 
    z_0 = c_{z, i} - s_i, &  z_1 = c_{z, i} + s_s
    \end{array}
\end{equation}

where $c_i$ is the full-scale centroid coordinates for the $i^{th}$ vertebra, and $s$ are the bounding box sizes. The subscripts for the sizes $s$ correspond to left ($l$), right ($r$), posterior ($p$), anterior ($a$), inferior ($i$), and superior ($s$). All six are necessary as the centroids defined here are vertebral body centers as the bounding box is not symmetric around the centroid as it includes posterior elements of the vertebra.

Bounding box sizes are regressed using a log Intersection-over-Union (IoU) loss function \cite{Tian2019}\cite{Tian2020}\cite{Yu2016a}:

\begin{equation}
    \begin{array}{c}

    L_{BB} = -\text{log}(IoU) \\
    L_{BB} = -\text{log}(\frac{b \, \cap \, \hat{b}} {b \, \cup \, \hat{b}})
    \end{array}
\end{equation}

where $\hat{b}$ is the predicted bounding box and $b$ is the ground truth. This loss was used as opposed to mean-absolute-error (MAE) or smooth-L1 as it allows for box sizes to be slightly modified if the centroid prediction is shifted (ie, larger left than right if the centroid prediction is slightly offset).

\subsection{Classification Branch}

\begin{figure}[h!]
    \centering
 	\includegraphics[width=0.7\linewidth]{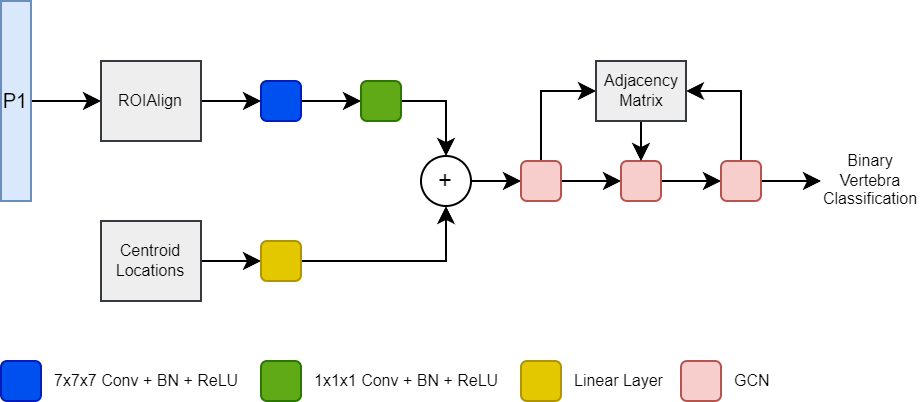}
	\caption{Detailed architecture of the classification branch of the VertDetect model. The \textit{P1} block is the feature map from the ResNet-50 + FPN backbone. The linear layer takes three input channels (for the x, y, and z coordinates) and outputs a feature map the same size as the output of the 1x1x1 convolution block.} 
    \label{class_branch}
\end{figure}

The purpose of the classification branch is to leverage the information between neighbouring vertebrae to improve overall classification and detection. 

Fig.~\ref{class_branch} shows the classification branch. A RoiAlign [32] layer, which is used to crop and resample features, first generates \textit{C} feature maps of size 7x7x7 from P1 cropping and resampling regions focused on the centroid locations. The resampled feature maps are then sent through a 7x7x7 followed by a 1x1x1 convolution, both with ReLU activation. Each cropped region loses any positional information it may have with its neighbours due to the cropping with the RoiAlign. Using the centroid locations predicted from the heatmaps the position of each region is encoded and combined with the resampled features as shown in \ref{class_branch}. The resulting features are then sent to three Graph Convolutional Network (GCN) layers to leverage the shared information between each vertebra. The result is a tensor with \textit{C}x1 logits that correspond to a vertebra being present in the scan or not. As the class of each vertebra is implicitly defined by the heatmap’s channel a binary classification is used to determine if that channel and predicted centroid correspond to a positive detection. A vertebra is positively detected if $\text{sigmoid}(q_i) > 0.5$ where  $q_i$ is the logit score for the $i^{th}$ vertebra of the classification branch output. The classification branch is trained using binary cross entropy, $L_{class}=BCE$,

\begin{equation}
    L_{class} = -\frac{1}{N} \sum_{i} y_i \text{log}(p_i) + (1 - y_i)\text{log}(1-p_i)    
\end{equation}

where $y_i$ is the binary value specifying if the $i^{th}$ vertebra is active, $p_i$ is the predicted probability vertebra $i$ being active, and $N$ is all possible vertebra. 

\subsection{Segmentation Branch}

The segmentation branch semantically segments positively detected vertebra. Before extracting the positively detected regions found in the previous branches, an unnormalized Gaussian heatmap is concatenated with the full resolution input image, as seen in \ref{mod_arch}. This Gaussian is centred about the predicted full-resolution centroid locations with a standard deviation of 4. As the vertebra segmentation step is for the full vertebra, bounding boxes will contain neighbouring vertebra due to the inclusion of posterior elements. The Gaussian heatmap ensures that the model focuses on the correct vertebra during semantic segmentation. 

Positively detected regions from both the P1 and the Gaussian input are extracted with a RoIAlign by cropping and resampling regions to 16x24x24 and 32x48x48, respectively. The convolved features originating from P1 are upsampled and concatenated with the features originating from the Gaussian-input image. The resulting feature map is sent through the final convolutional layers as shown in Fig.~\ref{mod_arch} to compute the segmentation predictions. The predicted segmentations are trained using binary cross-entropy loss, $L_{seg}=BCE$. Similar to Yi et al. \cite{Yi2021}, all convolution and transpose convolution operations (except for the final prediction layer) in the segmentation branch use instance normalization. As the classification and detection of each vertebra happens earlier in the model, the sole objective of this branch is to carry out semantic segmentation. Therefore, instance normalization allows for each cropped/resampled region from the RoIAlign to be treated independently. 

\subsection{Loss Functions}

The loss functions for training are the sum of all the loss functions previously discussed and an additional loss function $L_{}dist$, which is the Euclidean distance between adjacent vertebrae normalized by the total heatmap image size. 

\begin{equation}
    L_{total} = L_{heat} + L_{offset} + L_{BB} + L_{class} + L_{dist}    
\end{equation}

\subsection{Label Ordering Adjustment}
\label{sec: post_process}

\begin{figure}[h!]
    \centering
	\includegraphics[width=0.8\linewidth]{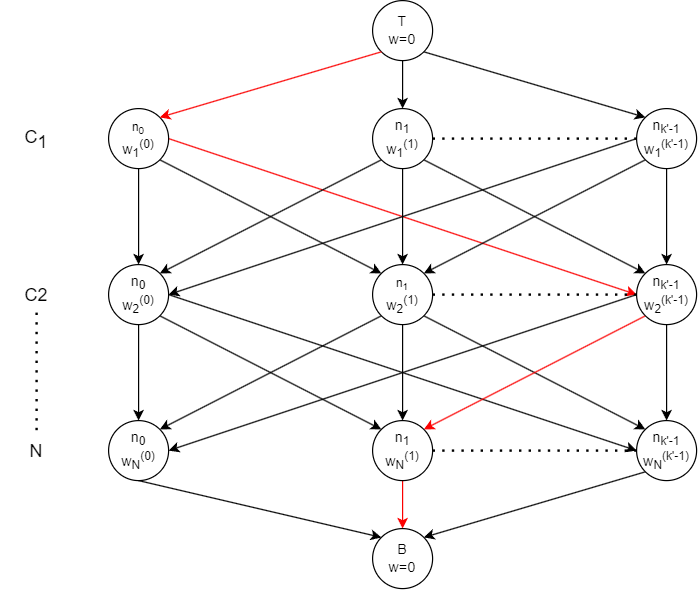}
	\caption{Example of the post-processing graph network. Each node represents a potential centroid location with the bottom value of each node corresponding to the weight value for that node. Each row of the graph corresponds to the potential locations for the same vertebra label. The \textit{T} and \textit{B} correspond to the top and bottom, respectively, and both have weights of zero. The red lines show an example of the path from T to B solving the graph and determining the nodes and therefore corresponding locations for that detection.} 
    \label{graph_net}
\end{figure}

As seen in Fig.~\ref{heatmap_slice} the predicted heatmaps have can have multiple local clusters and these local clusters can incorrectly correspond to the neighbouring vertebrae. To address the local maxima that can occur in the heatmap predictions, a post-processing method is used to determine which maxima from the local clusters are correct. A non-maximum suppression (NMS) is first used through a max-pooling layer to select the top $k$ candidates from each channel of the heatmap predictions. These k candidates are then filtered by euclidean distances to ensure no neighbours from the same local cluster exist resulting in $k^\prime$ candidates for each heatmap channel. The logits for each $k^\prime$ candidate from the heatmap predictions are then averaged with the logits from the classification branch for each corresponding vertebrae to scale each based on the probability of that particular vertebra existing in the scan. The resulting $k^\prime$ candidates are constructed in a graph as seen in \ref{graph_net} based on two rules. The first rule is that the axial position of the node above must be greater than the node below to ensure the correct vertebrae ordering is enforced. The second is that the Euclidean distance between any two connected nodes must be greater than 3 voxels to ensure that no two nodes from the same physical locations are used. The weights for each node are taken as the averaged logits. The centroid location of each vertebra is then determined by solving the graph from \textit{T} (top) to \textit{B} (bottom) by determining the longest path and therefore the path with the highest sum of averaged logits. \ref{graph_net} shows an example of this with the red lines indicating the solved path from the T to B with the resulting centroids corresponding to $C_1^{(0)} \rightarrow C_2^{(k^\prime-1)} \rightarrow ... \rightarrow C_N^{(1)}$. This process is also performed from T to B and from B to T with the path with the largest sum taken as the correct path.

\section{Evaluation Metrics}
All experiments used Dice Similarity Coefficient (DSC) as the metric for comparison. The DSC of a vertebra is also based on its class prediction; if two vertebrae have similar semantic segmentations but mismatched vertebral labels, the resulting DSC is 0 for that sample.

\section{Implementation}

VertDetect was trained on the merged 2019 and 2020 VerSe \cite{Sekuboyina2021} dataset, using the updated subject-based data structure, which includes anatomical labels of 26 vertebral levels (C1 to L5 as well as transitional T13 and L6 vertebrae). This consisted of 141 training, 120 validation, and 113 testing samples from the VerSe dataset. The validation and testing datasets here were the original testing and hidden datasets used in the challenge. 

All images were resampled to 1.75 mm$^3$ voxel spacing and either padded or cropped to 128x128x384 voxels based on the location of the ground truth segmentations to ensure all labels were in the image. Affine augmentation (translation, rotation, scaling, and flipping) and elastic deformation were used during training. 

For the detection branch $\alpha=2$ and $\beta=4$ for $L_{focal}$ and $\epsilon^\prime=100$ for the epoch threshold to transition between the MSE and variant focal loss. The ground truth Gaussian heatmaps used a $\sigma$ of 3.0 initially, and further changed to 2.0 after the self-initialization. A batch size of 4 was used and split across the 4 GPUs using distributed parallelization, resulting in 1 sample per GPU. AdamW optimizer was used with a static learning rate of 1e-4. The base filter size for the ResNet-50 backbone was 64 with \textit{d}=256 in the FPN.

As the heatmap output is critical for both the detection, graph classification and segmentation branches, the model was first trained with only the heatmap output for 500 epochs. This is referred to as the self-initialization. After the self-initialization, all outputs were predicted and all loss functions were used. During the first 100 epochs \textit{after} the self-initialization (between epochs 501 and 600 from the overall start) the model was trained using the ground truth bounding boxes for the segmentation task. This was done so to ensure the model learned to accurately segment vertebrae without being negatively affected by inaccurate bounding box predictions during the early stages of training. After this epoch threshold and when $L_{heat}<1.0$ the model transitioned to using the predicted bounding boxes. This continued until model completion. This threshold for $L_{heat}$ is a pre-determined hyperparameter of the model. Overall, the model was trained for 1500 epochs. 

The training was done using the Digital Research Alliance of Canada Mist server \cite{ComputeCanada2022a} which consists of 32-cores (128-threads) of IBM Power9 CPUs, 256 GB of memory and 4 Nvidia Tesla V100 32GB GPUs. All training was done using PyTorch 1.8.1 and utilized the build-in mixed-precision tools to reduce memory during training. SimpleITK 1.2.0 was used in both data-loading and augmentation.

To assess the impact of the graph classification and the self-initialization components, an ablation study was performed in which VertDetect was compared to the model without using the classification branch (VertDetect \textit{w/o GCN}) and without the self-initialization (VertDetect \textit{w/o self-init}). It was also tested with and without using the Euclidean distance loss, $L_{dist}$. For the models without the classification branch, positive samples were detected when the maxima of a heatmap were greater than 0.5 for each heatmap channel and post-processing was done using only the local heatmap maxima in the graph calculation. Models were also trained using only the variant focal loss for the heatmap (VertDetect \textit{focal only}). The epoch which resulted in the greatest DSC on the validation data was then used on the testing data.

\section{Results}

\subsection{Ablation Study}

\begin{table*}[!h]

\caption{Average DSC for the validation and testing for the VerSe challenge datasets for the different ablation study models. Values presented are for average DSC and 95\% CI in brackets following the average. PP indicated ``post-processing" and ``w/o" indicates ``without" and ``w/" indicates ``with". GCN refers to the Graph Convolution Network block, dist refers to the Euclidean distance loss, and NC is no convergence. Bold values show the greatest average DSC in each column.}
% \label{results_ablation_2019}
\begin{subtable}[!h]{\linewidth}
\caption{VerSe 2019 challenge.}

\resizebox{\textwidth}{!}
{

\begin{tabular}{|l|cc|cc|}
\hline
 & \multicolumn{2}{c|}{Validation} & \multicolumn{2}{c|}{Test} \\ \hline
Model & \multicolumn{1}{c|}{w/o PP} & w/ PP & \multicolumn{1}{c|}{w/o PP} & w/ PP \\ \hline
VertDetect focal only w/o GCN & \multicolumn{1}{c|}{0.875 (0.801-0.908)} & 0.872 (0.803-0.907) & \multicolumn{1}{c|}{0.869 (0.804-0.905)} & 0.871 (0.806-0.905) \\ \hline
VertDetect focal only & \multicolumn{1}{c|}{0.87 (0.813-0.903)} & 0.866 (0.804-0.899) & \multicolumn{1}{c|}{\textbf{0.886} (0.841-0.911)} & \textbf{0.882} (0.835-0.909) \\ \hline
VertDetect w/o dist & \multicolumn{1}{c|}{0.89 (0.847-0.915)} & 0.874 (0.79-0.908) & \multicolumn{1}{c|}{0.862 (0.799-0.899)} & 0.838 (0.745-0.888) \\ \hline
VertDetect w/o GCN + dist & \multicolumn{1}{c|}{0.855 (0.792-0.892)} & 0.862 (0.812-0.892) & \multicolumn{1}{c|}{0.87 (0.789-0.902)} & 0.872 (0.781-0.903) \\ \hline
VertDetect w/o GCN & \multicolumn{1}{c|}{\textbf{0.89} (0.851-0.911)} & \textbf{0.883} (0.843-0.906) & \multicolumn{1}{c|}{0.862 (0.781-0.898)} & 0.859 (0.781-0.896) \\ \hline
VertDetect & \multicolumn{1}{c|}{0.877 (0.826-0.905)} & 0.873 (0.824-0.902) & \multicolumn{1}{c|}{0.869 (0.795-0.905)} & 0.862 (0.785-0.902) \\ \hline
VertDetect w/o self-init + GCN + dist & \multicolumn{1}{c|}{0.84 (0.747-0.888)} & 0.825 (0.728-0.879) & \multicolumn{1}{c|}{0.859 (0.774-0.895)} & 0.867 (0.778-0.902) \\ \hline
VertDetect w/o self-init + GCN & \multicolumn{1}{c|}{0.839 (0.746-0.884)} & 0.836 (0.75-0.884) & \multicolumn{1}{c|}{0.834 (0.752-0.877)} & 0.849 (0.762-0.888) \\ \hline
VertDetect w/o self-init & \multicolumn{1}{c|}{0.846 (0.786-0.879)} & 0.852 (0.786-0.887) & \multicolumn{1}{c|}{0.846 (0.797-0.878)} & 0.871 (0.823-0.896) \\ \hline
VertDetect MSE only & \multicolumn{1}{c|}{NC} & NC & \multicolumn{1}{c|}{NC} & NC \\ \hline
VertDetect MSE only w/o GCN & \multicolumn{1}{c|}{NC} & NC & \multicolumn{1}{c|}{NC} & NC \\ \hline
\end{tabular}

}

\label{results_ablation_2019}
\end{subtable}

\bigskip

\begin{subtable}[!h]{\linewidth}
% \caption{Average DSC for the validation and testing for the VerSe 2020 challenge datasets for the different ablation study models. PP indicated ``post-processing" and ``w/o" indicates ``without" and ``w/" indicates ``with". GCN refers to the Graph Convolution Network block, dist refers to the Euclidean distance loss, and NC is no convergence.}
\caption{VerSe 2020 challenge.}
\label{results_ablation_2020}
\resizebox{\textwidth}{!}
{

\begin{tabular}{|l|cc|cc|}
\hline
 & \multicolumn{2}{c|}{Validation} & \multicolumn{2}{c|}{Test} \\ \hline
Model & \multicolumn{1}{c|}{w/o PP} & w/ PP & \multicolumn{1}{c|}{w/o PP} & w/ PP \\ \hline
VertDetect focal only w/o GCN & \multicolumn{1}{c|}{0.845 (0.809-0.872)} & 0.856 (0.821-0.881) & \multicolumn{1}{c|}{0.85 (0.815-0.876)} & 0.851 (0.814-0.877) \\ \hline
VertDetect focal only & \multicolumn{1}{c|}{0.845 (0.81-0.87)} & 0.84 (0.801-0.868) & \multicolumn{1}{c|}{0.855 (0.823-0.878)} & 0.856 (0.825-0.879) \\ \hline
VertDetect w/o dist & \multicolumn{1}{c|}{\textbf{0.861} (0.824-0.886)} & 0.855 (0.812-0.882) & \multicolumn{1}{c|}{\textbf{0.862} (0.826-0.887)} & 0.852 (0.807-0.881) \\ \hline
VertDetect w/o GCN + dist & \multicolumn{1}{c|}{0.84 (0.803-0.865)} & 0.855 (0.82-0.879) & \multicolumn{1}{c|}{0.852 (0.814-0.877)} & 0.862 (0.824-0.886) \\ \hline
VertDetect w/o GCN & \multicolumn{1}{c|}{0.852 (0.819-0.875)} & 0.86 (0.823-0.884) & \multicolumn{1}{c|}{0.86 (0.823-0.883)} & \textbf{0.869} (0.832-0.891) \\ \hline
VertDetect & \multicolumn{1}{c|}{0.859 (0.826-0.882)} & \textbf{0.868} (0.834-0.89) & \multicolumn{1}{c|}{0.849 (0.808-0.876)} & 0.849 (0.805-0.878) \\ \hline
VertDetect w/o self-init + GCN + dist & \multicolumn{1}{c|}{0.836 (0.792-0.865)} & 0.836 (0.792-0.866) & \multicolumn{1}{c|}{0.849 (0.809-0.875)} & 0.856 (0.816-0.882) \\ \hline
VertDetect w/o self-init + GCN & \multicolumn{1}{c|}{0.818 (0.781-0.848)} & 0.815 (0.77-0.848) & \multicolumn{1}{c|}{0.819 (0.782-0.846)} & 0.834 (0.793-0.862) \\ \hline
VertDetect w/o self-init & \multicolumn{1}{c|}{0.832 (0.795-0.858)} & 0.838 (0.796-0.868) & \multicolumn{1}{c|}{0.84 (0.804-0.865)} & 0.848 (0.808-0.874) \\ \hline
VertDetect MSE only & \multicolumn{1}{c|}{NC} & NC & \multicolumn{1}{c|}{NC} & NC \\ \hline
VertDetect MSE only w/o GCN & \multicolumn{1}{c|}{NC} & NC & \multicolumn{1}{c|}{NC} & NC \\ \hline
\end{tabular}

}

\end{subtable}

\end{table*}

Tables \ref{results_ablation_2019} and \ref{results_ablation_2020} show the validation and testing results for the VerSe 2019 and 2020 data-sets, respectively, for multiple VertDetect configurations, and with arbitrary fields-of-view. The variant focal loss was shown to be necessary for training as models trained with MSE alone were not able to converge. However, the linear scheduling using both the variant focal loss and MSE did not have a significant effect. The self-initialization showed the benefit of the localization of the variant focal loss and its ability to improve model convergence. The results also show that the post-processing either matched or improved DSC for all models except one (VertDetect w/o dist) indicating its usefulness.  The GCN and the euclidean distance loss did not improve model accuracy but did improve model stability, which is discussed later.

\subsection{Results Comparison}

\begin{table*}[!h]
    \centering
        \caption{Current state-of-the-art models for the VerSe segmentation challenge, both for the 2019 and 2020 years. Included is also a brief description of the architectures/designs used by the authors.}
\resizebox{1\textwidth}{!}{

    % \begin{tabular}{|c|c|c|c|c|c|c|c|c}
    \begin{tabular}{|p{2cm}|p{6cm}|c|c|c|c|}
    % \begin{tabularx}{\linewidth}{X*{5}{>{\centering\arraybackslash}X}}

\hline
    
Author(s) & Model Design & \multicolumn{2}{c|}{VerSe 2019} & \multicolumn{2}{c|}{VerSe 2020}  \\

\hline

\hfill & \hfill & Public Test & Hidden Test & Public Test & Hidden Test \\

\hline

Payer C. & Multi-stage; classification followed by segmentation & 0.909 & 0.898 & 0.916 & 0.897 \\
\hline

Chen D. & Multi-stage; segmentation to classification & --- & --- & 0.917 & 0.912 \\ 
\hline

Lessmann N. & Iterative 3D U-Net with classification leg & 0.851 & 0.858 & --- & --- \\ 
\hline 

Chem M. &  Multi-stage; 3D U-Net performs segmentation followed by R-CNN for labelling & 0.930 & 0.826 & --- & --- \\ 
\hline

Zhang A. & Multi-stage; 3D V-Net to predict candidates followed by network for segmentation and post-processing for labelling  & --- & --- & 0.888 & 0.894 \\
\hline

Yeah T. & Multi-stage; 3D U-Net used for localization at low resolution followed by second 3D U-Net for segmentation at higher resolution & --- & --- & 0.889 & 0.879 \\
\hline

Xiangshang Z. & Multi-stage; Btrfly-Net \cite{Sekuboyina2018a} for key-point detection followed by nnU-Net for segmentation \cite{Isensee2018}  & --- & --- & 0.836 & 0.851 \\
\hline

Tao et al. & Multi-stage; Iterative 3D transformer model to detect vertebrae followed by encoder-decoder for segmentation & 0.911 & 0.901 & --- & --- \\
\hline

You et al. & Iterative 3D transformer with global information transformer (ignored T13 in dataset)& 0.864 & 0.865 & 0.845 & 0.868 \\
\hline

\textit{VertDetect} & Single stage detection & 0.883 & 0.882 & 0.868 & 0.869 \\
\hline

    \end{tabular}}

    \label{results_comparison}
\end{table*}

Table \ref{results_comparison} shows how VertDetect compares to other state-of-the-art models for vertebral instance segmentation for the 2019 and 2020 VerSe validation and testing data. VertDetect shows comparable performance (0.0178 to 0.0534 DSC difference) to the other state-of-the-art models but achieves greater performance than other single end-to-end 3D models.

\section{Discussion}

\subsection{Model Performance}

VertDetect is able to achieve its performance on-par with other existing models for vertebral instance segmentation in a single end-to-end model utilizing the full 3D CT scan. This design is more efficient than those using multiple cascading models. Furthermore, the feature space captured by VertDetect is derived from the full spine image rather than cropped patches; this additional contextual information may be of value for downstream tasks where the geometry of the whole spine is important. 

Euclidean distance loss was not found to show a clear benefit (Tables \ref{results_ablation_2019} and \ref{results_ablation_2020}). The magnitude of the distance loss was on average 1, which is larger in magnitude than the other losses. This increases the total loss sum and could cause difficulties during backpropagation. Further experimentation would be needed to assess the effect of re-weighting the contribution of the distance loss relative to the other loss terms.

Tables \ref{results_ablation_2019} and \ref{results_ablation_2020} It was also found that that the GCN layer does not improve overall model accuracy but the reason for this is unclear. The GCN takes a feature representation from the model backbone and uses RoiAlign and convolution layers to generate the node features for the graph. It is possible that the initial feature maps used in this layer are not appropriate. The model uses high-resolution features but those that are further away from the heatmap classification stage. This was done to avoid the potential spareness of the feature maps that could arise due to the variant focal loss, but it is possible that the implementation of this could be better optimized by using more advanced GCN layer designs, different feature representations, or objectives like solving for the adjacency matrix. 

The most direct comparison of VertDetect can be found in the model by You et al.~\cite{You2022,You2023} which utlizes a full CT scan without relying on iterative or cascading modelling approaches. While this model does rely on cropped patches from whole 3D volumes for inputs to their vision transformer \cite{Dosovitskiy2020}, they also capture whole scan information (a downsampled version of the CT volume) using a second transformer. The features from the are combined with those from the transformer with the cropped patch input. Table~\ref{results_comparison} shows that VertDetect was able to achieve a marginally better performance than \cite{You2022,You2023} with the inclusion of the T13 vertebra (You et al. did not consider T13). Removal of the T13 vertebra from the analysis of VertDetect led to improved DSC in the VerSe 2020 public and hidden test sets of 0.872 (95\% CI, 0.837-0.893) and 0.874 (95\% CI, 0.837-0.895), respectively.

Similarly to the other models outlined in the VerSe challenge paper \cite{Sekuboyina2021}, detection of T13 transitional vertebrae poses difficulties. The difficulty in T13 detection seems to be mainly caused by the low sampling frequency (2 training, 2 validation, 2 testing) in the challenge overall. Challenges also arise due to the anatomy surrounding the T13 vertebrae. In manual labelling the presence of small, floating ribs can be used to distinguish T13 vertebrae. However, specifically for VertDetect, these ribs may be too small for the model to properly distinguish, leading to misclassification as L1. If the remaining lumbar region is visible beyond this point, VertDetect will further classify L5 as L6. VertDetect is able to determine that a transitional vertebra is present in the scan, but it has a difficult time properly distinguishing which transitional vertebra is included (T13 or L6). It is possible that oversampling the samples with T13 vertebra during training, or adding additional T13 and L6 cases to the training set, could help to overcome this problem. 

\begin{figure}[h!]
	\includegraphics[width=\linewidth]{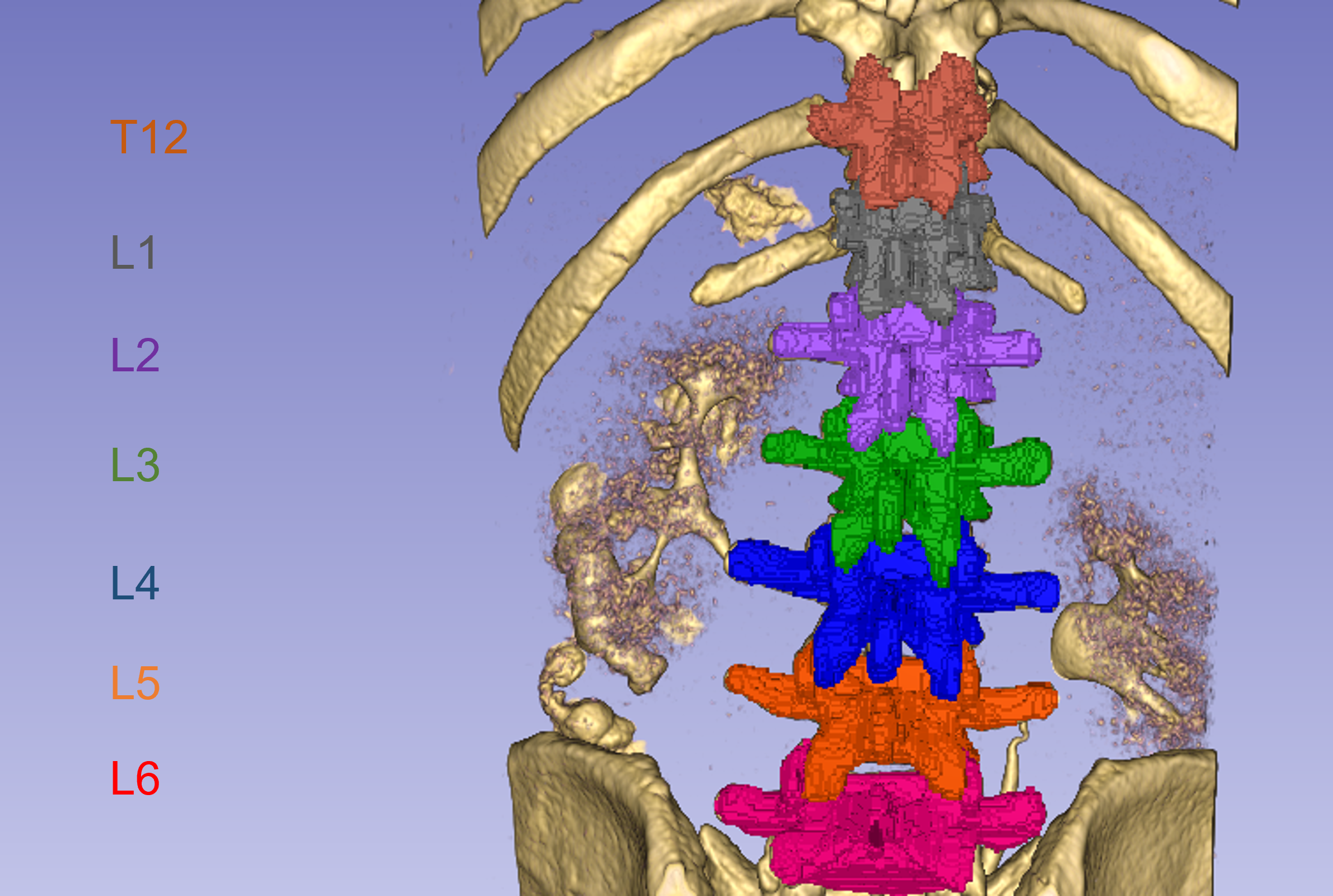}
	\caption{3D rendering of sample from VerSe 2020 validation dataset highlighting validation potentially mislabeled lumbar vertebra. Vertebra labels and segmentations are from the ground truth.} 
    \label{gt_sample_707}
\end{figure}

In the validation (public test) dataset, there is a single sample that was not adequately identified in \textit{all} experiments and model development. This sample is of the lower lumbar region, with ground truth segmentations ranging from T12 to L6, and a 3D rendering of this sample with ground truth segmentations can be seen in Figure~\ref{gt_sample_707}. During all experiments with VertDetect, the L6 is predicted as an L5. The 3D rendering shows that the ground truth L1 has some protrusions. The protrusions on this vertebra may be floating ribs and as such this vertebra may be thoracic and not lumbar. If this is the case, then the model's prediction of L5 is accurate and the ground truth labels have errors. Due to the post-processing in section~\ref{sec: post_process}, this mislabel between L5 and L6 would cause all other vertebrae in this scan to be mislabeled. 

\subsection{Model Stability}

\begin{figure}[h!]
	% \includesvg[width=\linewidth]{Figures/validation_dsc_training_curve}
 	\includesvg[width=\linewidth]{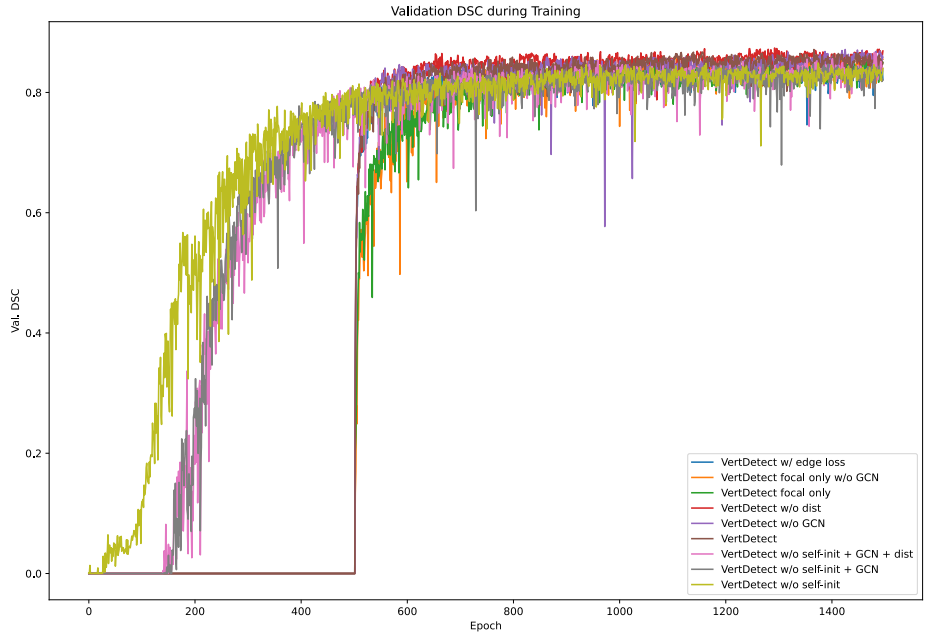}
	\caption{Validation DSC during training. Validation DSC was done without using the post-processing to save time during training. The spike at epoch 500 is due to the self-initialization as the segmentation branch is not active for those models until that epoch.} 
    \label{val_curve}
\end{figure}

\begin{figure}[h!]
	% \includesvg[width=\linewidth]{Figures/validation_dsc_training_curve_blown_up}
 	\includesvg[width=\linewidth]{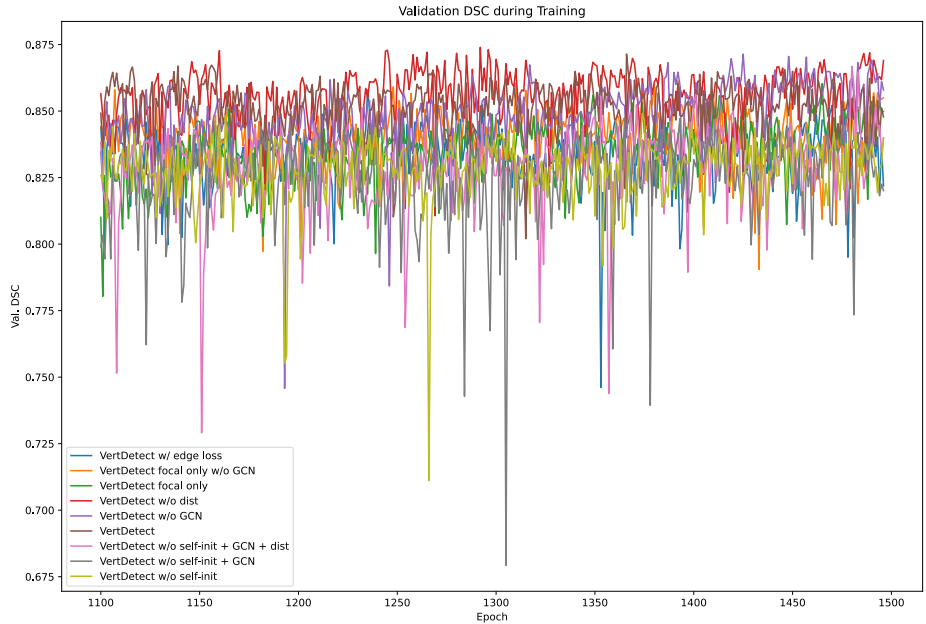}
	\caption{Magnified portion of Fig.~\ref{val_curve} from epoch 1100 to better highlight the variability in the validation DSC between different experiments.} 
    \label{val_curve_blown_up}
\end{figure}
 
Model stability is a significant factor that is often left out when discussing CNN performance. Due to the complexity of VertDetect and the multiple loss functions, model stability is an important aspect to consider. Figures \ref{val_curve} and \ref{val_curve_blown_up} show the DSC of the validation set during training (without using post-processing) of all epochs, and the final 400 epochs, respectively. The flat line of zero validation DSC in Figure \ref{val_curve} for some models is due to the self-initialization for the first 500 epochs. Figures \ref{val_curve} and \ref{val_curve_blown_up} show that the model with self-initialization, GCN and distance losses has the smallest fluctuation in validation DSC. Decreases in the validation DSC are mainly caused by problems in the heatmap's ability to properly identify the correct vertebra. The distance losses and GCN classification were designed to assist with the heatmap centroid predictions and the implicit vertebra labelling, respectively. The self-initialization is also necessary to help with the initial convergence of the variant focal loss prior to the other parts of the model being enabled. As shown in Figure \ref{val_curve_blown_up}, the models without the self-initialization have spikes of significantly decreasing validation DSC, indicating model instability.  

Figures \ref{val_curve} and \ref{val_curve_blown_up}, and Tables \ref{results_ablation_2019} and \ref{results_ablation_2020}, show that combining the variant focal loss and MSE does not provide any benefit compared to using the variant focal loss only. The heatmap predictions are very sparse by design and this proved to be too difficult for model convergence when training with the other losses. This led to the combined MSE and variant focal loss. However, with the self-initialization, the variant focal loss is not competing with other losses so convergence is easier. This is important as a 2x downsample was used. If no downsampling was used the sparseness of the heatmap prediction is increased, the combined loss may be more impactful, but this requires more experimentation and compute power. 

From the existing models outlined in Table~\ref{results_comparison}, only Payer C. \cite{Payer2020} has both the model and dataloader publicly available on their GitHub. This model was run with both the original VerSe 2020 data and the updated 2020 data used here. Using the validation data from the original and updated 2020 datasets the model achieved a DSC of 0.883 and 0.803, respectively. The result from the original data is on par with the reported results from the VerSe challenge and the results shown here. However, when trained and validated on the updated dataset, the resulting DSC is significantly lower. The purpose of this is not to speculate as to the reason for the different DSC values, but to highlight the importance of model stability.

\subsection{Heatmap Local Clusters}

One complication in the heatmap predictions is the possibility of local clusters as seen in Fig.~\ref{heatmap_slice}. Fig.~\ref{heatmap_slice} shows two small clusters corresponding to the correct vertebra and the adjacent vertebra below. It is possible that the maximum intensity in the heatmap prediction aligns with the incorrect adjacent vertebra. A 2x downsampling (compared to the 4x downsampling used in \cite{Yi2021}) showed superior separation of the local clusters, however, the local clusters themselves seem to be unavoidable. The cluster with the maximum intensity is used to estimate the centroid location, but when multiple clusters are present this can lead to errors. This seemed to be an experimental artifact of the variant focal loss, however, the variant focal loss demonstrated the greatest performance in centroid prediction as the MSE loss failed to converge.

To address the local clustering issue in the heatmap predictions, a post-processing method was developed in section~\ref{sec: post_process}. Tables \ref{results_ablation_2019} and \ref{results_ablation_2020} show the performance increase when using the post-processing. Since multiple centroids cannot all point to the same vertebra, the post-processing considers all possible local clusters for a scan and tries to determine the correct ordering. There are two current complications with this approach. The first is that it only considers centroid probabilities rather than also considering the Euclidean distance between vertebrae. This was explored but a strong solution that combined both probability and distance values was not found (and using probability values only showed better performance). Second, the post-processing method relies on strong performance from the model as is clear from Tables \ref{results_ablation_2019} and \ref{results_ablation_2020}. The post-processing goes top-down (inferior direction) and uses the first predicted vertebral label to determine the subsequent vertebral labels. If this initial vertebra label is wrong then the post-processing will fail as all subsequent labels are incorrect.

\subsection{Future Improvements}

The VertDetect model is large leaving little GPU memory headroom available for improvements. The model was trained on four Nvidia V100 GPUs with 32 GB of memory each. As larger GPUs become available however, there are some features/changes that could be considered to improve performance. The input images required significant downsampling, especially in the axial direction, to overcome the memory-bottleneck issues. Less downsampling in the axial direction could improve the separation of neighbouring vertebrae with more image sharpness and less interpolation leading to better detection capabilities. This is supported by improvements in performance achieved by changing from 4-times downsampling to 2-times. It is possible that further improvements can be gained through a zero-downsampled heatmap, but will require significantly more GPU memory.  

The graph network in the classification block uses features from the ResNet-50 + FPN backbone, but ideally, it could use the heatmap predictions themselves. This was attempted in early iterations of this work with a framework similar to Yang et al. \cite{Yang2017}, but proved too computationally intensive. A stronger connection between the heatmaps and the classification branch would be beneficial and could be achieved using a message-passing framework based on the heatmap predictions.

All models shown in Tables~\ref{results_ablation_2019} and \ref{results_ablation_2020} took 7 days to train. However, it is possible that the models may still benefit from further training time should suitable computing hardware become available.

\section{Conclusion}

The task of vertebral instance segmentation of 3D CT scans is challenging due to the complex 3D shape of individual vertebrae and the similarity in the shape of neighbouring vertebrae. However, the automation of this task is highly desirable for clinical use to reduce the workload of radiologists, surgeons and other medical professionals in downstream tasks for diagnoses, navigation, and planning. VertDetect, a model that can accurately perform this instance segmentation task in 3D utilizing a single end-to-end structure. This allows 3D features of the spine and vertebral levels to be used in the detection and segmentation stages, and better utilizes the known structure of the spine in final segmentation predictions.

\section{Acknowledgements}
High-performance computing resources were provided by the Digital Research Alliance of Canada, SciNet, and Smart Computing for Innovation. This work was funded by the Canadian Institutes of Health Research (CIHR), Varian Medical Systems, INOVAIT through the Government of Canada's Strategic Innovation Fund, and Feldberg Chair for Spinal Research.

\bibliographystyle{model2-names_rev2.bst}
\bibliography{UofT_PhD}

\end{document}